\documentclass[eqsecnum,preprint]{elsarticle}
\usepackage{amsmath}
\usepackage{epsfig,epsf}
\usepackage{graphicx}
\usepackage{verbatim}
\usepackage{epstopdf}

\def\delta{{\mathbf \Delta}}

\newcommand{\beq}{\begin{eqnarray}}
\newcommand{\eeq}{\end{eqnarray}}
\newcommand{\be}{\begin{eqnarray*}}
\newcommand{\ee}{\end{eqnarray*}}

\def\be{\begin{equation}}
\def\ee{\end{equation}}
\def\bea{\begin{eqnarray}}
\def\eea{\end{eqnarray}}

\begin{document}


\voffset1.5cm

\title{Kinematic biases on centrality selection of jet events in pPb collisions at the LHC}

\author{N\'estor Armesto}

\address{Departamento de F\'isica de Part\'iculas and IGFAE,\\
Universidade de Santiago de Compostela,
E-15706 Santiago de Compostela,
Galicia-Spain}

\ead{nestor.armesto@usc.es}

\author{Do$\breve{\rm g}$a Can G\"ulhan}

\address{Laboratory for Nuclear Science and Department of Physics,\\
Massachusetts Institute of Technology
(MIT), Cambridge, MA 02139 USA}

\ead{dgulhan@mit.edu}

\author{Jos\'e Guilherme Milhano}

\address{CENTRA, Instituto Superior T\'ecnico,\\
Universidade de Lisboa, Av. Rovisco Pais, P-1049-001
Lisboa, Portugal\\
and\\
Physics Department, Theory Unit, CERN, CH-1211 Gen\`eve 23, Switzerland}

\ead{guilherme.milhano@tecnico.ulisboa.pt}



\cortext[thanks]{Preprint number: CERN-PH-TH-2015-021.}

\begin{abstract}

Centrality selection has been observed to have a large effect on jet observables in $p$Pb  collisions at the Large Hadron Collider, stronger than that predicted by the nuclear modification of parton densities. We study to which extent simple considerations of energy-momentum conservation between the hard process and the underlying event affect jets observables in such collisions. 
We develop a simplistic approach that considers first the production of jets in a $pp$ collision as described by PYTHIA.
From each $pp$ collision, the value of the energy of the parton from the proton participating in the hard scattering is extracted. Then, the underlying event is generated simulating a $p$Pb collision through HIJING, but with the energy of the proton decreased according to the value extracted in the previous step, and both collisions are superimposed.
This model is able to capture the bulk of the centrality effect for central to semicentral collisions, for the two available sets of data: dijets from the CMS Collaboration and single jets from the ATLAS Collaboration. As expected, the model fails for peripheral collisions where very few nucleons from Pb participate.

\end{abstract}

\maketitle

\section{Introduction}
\label{intro}

The classification of events according to some measurement of their activity (energy, multiplicity, etc.)  in some region of phase space, what is commonly referred to as  centrality selection, is extensively used in the study of nuclear collisions. 
Naively one expects that to events within a given centrality class correspond similar conditions, say of energy density or temperature. The dependence of specific observables on centrality should then yield information on their sensitivity to the properties of the system (energy density, temperature, etc.). Further, centrality should provide a link, albeit in a model dependent way, to theoretically well-defined quantities such as the impact parameter of the collision.


The usefulness of centrality selection relies, however, on how robustly it can be defined. Specifically, whether the centrality selection (and its link to some theoretical quantity as the impact parameter) depends (i) on the centrality criterium;  and (ii) on the observable under consideration i.e. whether there is a correlation between the characteristics of the events under consideration, say the requirement of presence of jets, and the centrality criterium.


While in non-peripheral collisions between heavy ions the classification of centrality classes appears to be robust, in proton-lead collisions at the Large Hadron Collider (LHC) the situation is far more problematic\footnote{The problem also exists in dAu collisions at the Relativistic Heavy Ion Collider (RHIC), see e.g. \cite{Adare:2013esx,STAR}.}. 
Whereas for both dijets \cite{Chatrchyan:2014hqa} and single jets \cite{ATLAS:2014cpa} the minimum bias results are well reproduced by standard pQCD with nuclear modification of parton densities \cite{Paukkunen:2014pha}, the centrality-selected results show a strong dependence on centrality that cannot be accommodated by the existing ideas on the impact parameter dependence of nuclear parton densities \cite{Armesto:2003fi,Frankfurt:2011cs,Helenius:2012wd}. There is an ongoing discussion on the definition of centrality in such asymmetric systems \cite{Adare:2013nff,Adam:2014qja}.

A numbers of factors can conceivably be at the origin of the observed effects. Fluctuations in the number of partipating nucleons are much larger in proton-nucleus than in nucleus-nucleus collisions and certainly confound the connection of centrality to collision impact parameter. Further, energy-momentum conservation poses significantly more stringent demands in the case of asymmetric collisions. Ultimately the problems can be traced back to our lack of understanding of the detailed microscopic dynamics underlying soft particle production in hadronic and nuclear collisions, and its coupling to hard subprocesses. These issues have been discussed in several recent works \cite{Martinez-Garcia:2014ada,Bzdak:2014rca,Alvioli:2014eda,Perepelitsa:2014yta}.


In this short note we address solely, and in a simplistic framework, the role of kinematic constraints on the energy-momentum of the proton. We ask to what extent the requirement of a hard subprocess restricts the soft particle production which will eventually determine the centrality of the event\footnote{Take for example the production of a dijet pair at pseudorapidity $\eta_{\rm dijet}=2$ with leading jet $p_{T}=120$ GeV. The minimum momentum fraction taken from the proton is then $x_p\sim 0.35$.}.

We focus on CMS dijet \cite{Chatrchyan:2014hqa} and ATLAS single-jet \cite{ATLAS:2014cpa} results in $p$Pb collisions at 5.02 TeV/nucleon. We find that our simplistic implementation leads to large effects in good agreement with data. Importantly the model fails to describe data when the implementation is evidently deficient, that is when energy-momentum constraints on the Pb nucleus are expected to play a role.  

The manuscript is organised as follows: in section \ref{model} we describe the model that implements energy-momentum conservation constraints on the proton. Then, in  sections \ref{cms} and \ref{atlas}, we employ the model to analyse the extent to which CMS results on dijets and ATLAS results on single jets, respectively, can be understood as resulting from kinematic considerations. Finally, in section \ref{conclu} we present our conclusions.


\section{Description of the model}
\label{model}

The model that we employ uses PYTHIA \cite{Sjostrand:2006za} for the hard scattering and HIJING \cite{Wang:1991hta,Gyulassy:1994ew} for the underlying event. 
Each event (our results are based on samples of $10^5$ for dijets and $2\cdot 10^6$ for single jet spectra) is generated according to the following procedure:

\begin{enumerate}
\item generate a $pp$ event in PYTHIA with no underlying event (for a proton beam of $E_p=4$ TeV and a Pb beam of $E_{\rm Pb}=1.58$ TeV/nucleon, $\sqrt{s_{NN}}=5.02$ TeV) with the required characteristics: jets or dijets within the experimental kinematic cuts, and extract the momentum fraction $x_p$ of the hard parton from the proton participating in the $2\to 2$ hard subprocess. The nuclear modification of parton densities was not taken into account because for this kinematics their effect, as discussed previously (see also \cite{Paukkunen:2014pha}), is much smaller than the one observed in data;
\item generate a minimum bias pPb event in HIJING, for a proton beam of $(1-x_P)E_p$ and  a Pb beam of $E_{\rm Pb}$, i.e. we reduce the proton energy for the underlying event such that $\sqrt{s_{NN}}=2\sqrt{(1-x_p) E_p E_{\rm Pb}}$;
\item shift both events (that are generated in their respective center of mass) to a common frame (the LHC laboratory frame);
\item the HIJING event, generated as minimum bias, is reweighted according to its impact parameter such that scaling in the number of collisions $N_{coll}$, expected for hard events, is fulfilled;
\item the PYTHIA and HIJING events are superimposed, resulting in a full pPb event with the weight given in the previous step.
\end{enumerate}

Note that the momentum fraction $x_{\rm Pb}$ of the parton from the Pb participating in the $2\to 2$ hard subprocess is unchanged. This is clearly a deficiency of the model, which should accordingly be expected to fail for small $N_{coll}$ (when the number of participating nucleons from the Pb is small) and for large $x_{\rm Pb}$ (for the Pb-going side in pseudorapidity).

\section{CMS dijet results}
\label{cms}

In this section we consider the dijet measurements by CMS \cite{Chatrchyan:2014hqa}. We generate hard events in PYTHIA with jets reconstructed
within $|\eta| < 3$ using the anti-$k_T$ sequential recombination algorithm \cite{Cacciari:2008gp,Cacciari:2011ma} with a
distance parameter of 0.3, considering events required to have a dijet with
a leading jet $p_{T,1} > 120$ GeV/c, a subleading jet $p_{T,2} > 30$ GeV/c, and an azimuthal distance between them $\Delta \phi_{1,2} > 2\pi/3$. Then, for the full PYTHIA+HIJING generated event, the transverse energy in the region  $4 < |\eta| < 5$ is rescaled, to match that measured by CMS, by a factor that accounts for detector resolution effects that are not corrected by CMS. It is this rescaled energy that is  used to classify events in centrality.

\begin{figure}[!ht]
 \centering
\includegraphics[scale=0.3]{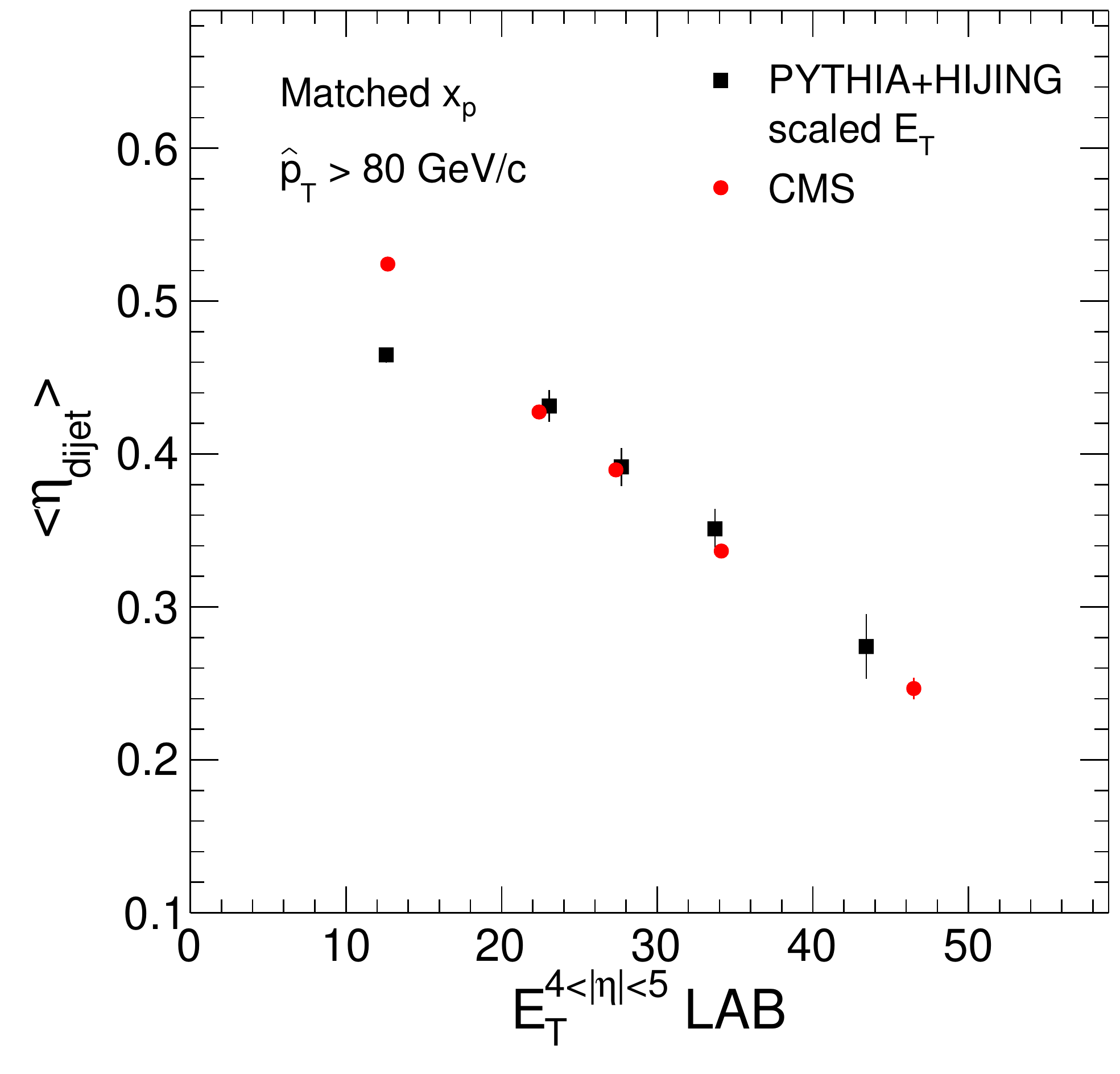}
\caption{Average dijet pseudorapidity ($\eta_{\rm dijet}=(\eta_{\rm 1}+\eta_{2})/2$) as a function of the total transverse energy deposition in the pseudorapidity range of $4 < |\eta| < 5$ for PYTHIA+HIJING events matched according to $x_{p}$ value (black squares) overlaid with data measured by CMS \cite{Chatrchyan:2014hqa} (red circles).}
\label{mean_dijeteta_ET}
\end{figure}

\begin{figure}[!ht]
 \centering
\includegraphics[scale=0.6]{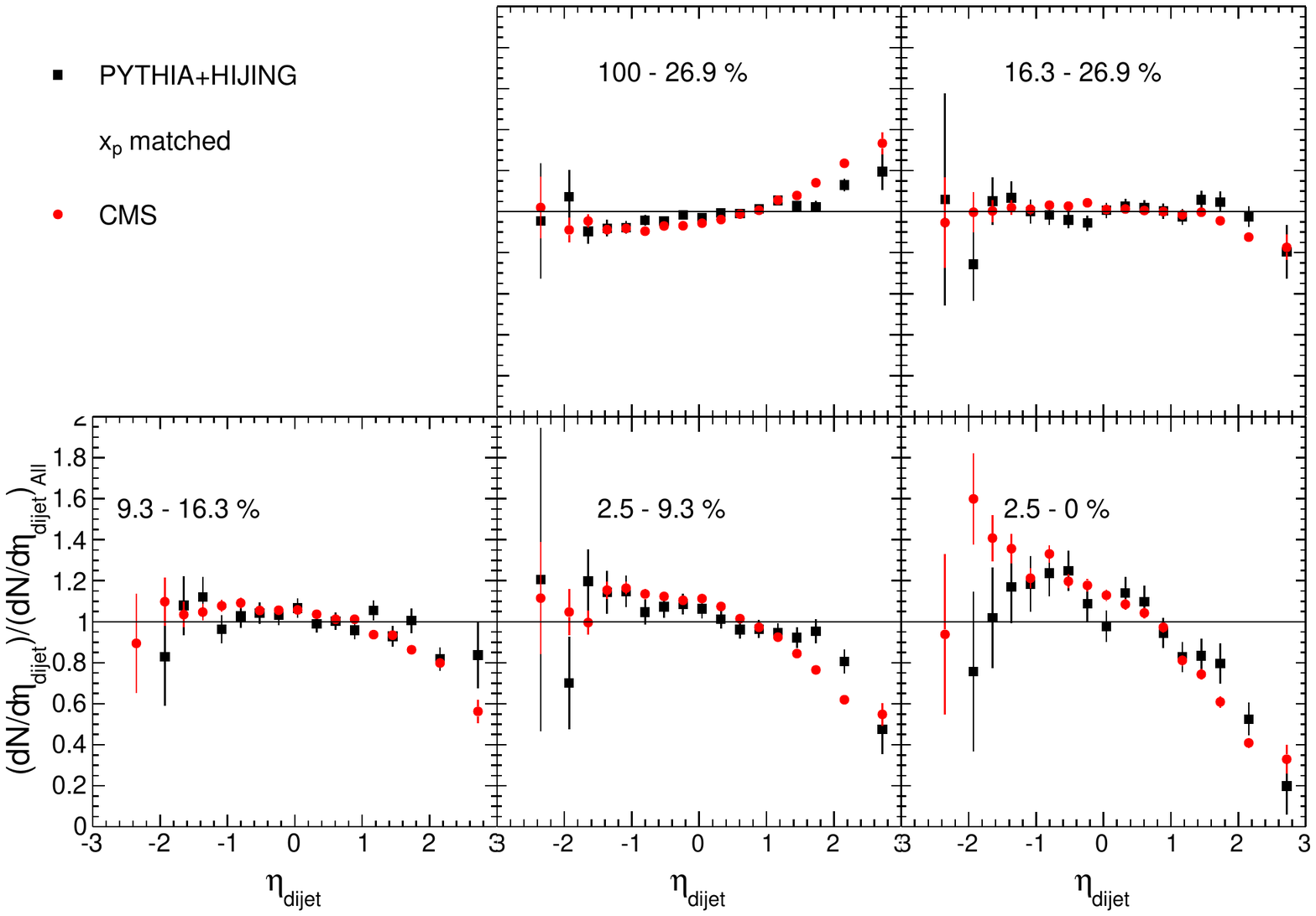}
\caption{Ratios of dijet pseudorapidity ($\eta_{\rm dijet}=(\eta_{1}+\eta_{2})/2$) distributions with a selection on total forward energy deposition (${\rm E}_{\rm T}^{[4 < |\eta| < 5]}$) to the dijet pseudorapidity distribution without any requirement on event activity. The calculation in PYTHIA+HIJING is shown by black squares and CMS data points \cite{Chatrchyan:2014hqa} are shown by red circles.}
\label{dijeteta_ratios}
\end{figure}

In figure \ref{mean_dijeteta_ET} we show the comparison of our results for the average dijet pseudorapidity ($\eta_{\rm dijet}=(\eta_{\rm 1}+\eta_{2})/2$) as a function of the total transverse energy deposition in the pseudorapidity range of $4 < |\eta| < 5$, compared with data measured by CMS \cite{Chatrchyan:2014hqa}. The agreement is very good except for the lowest transverse energies, corresponding to peripheral collisions where the model is not expected to work as discussed in the previous section. In figure \ref{dijeteta_ratios} we show the ratio of the dijet pseudorapidity distribution for events in a a given centrality class over the minimum bias distribution, compared to the CMS data. Again, a good overall agreement is found.

\begin{figure}[!ht]
 \centering
\resizebox{0.5\textwidth}{!}{\includegraphics{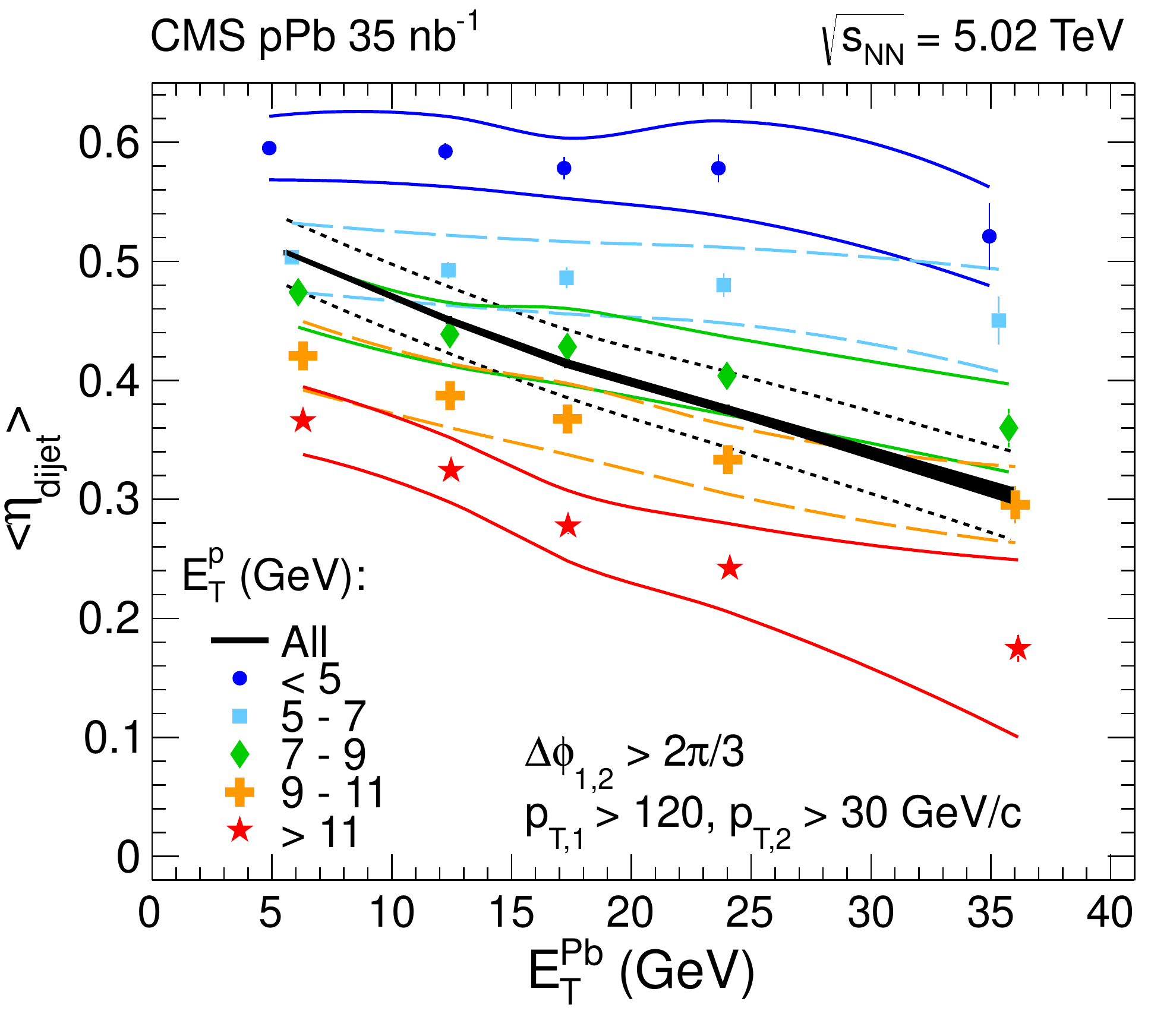}}
\resizebox{0.43\textwidth}{!}{\includegraphics{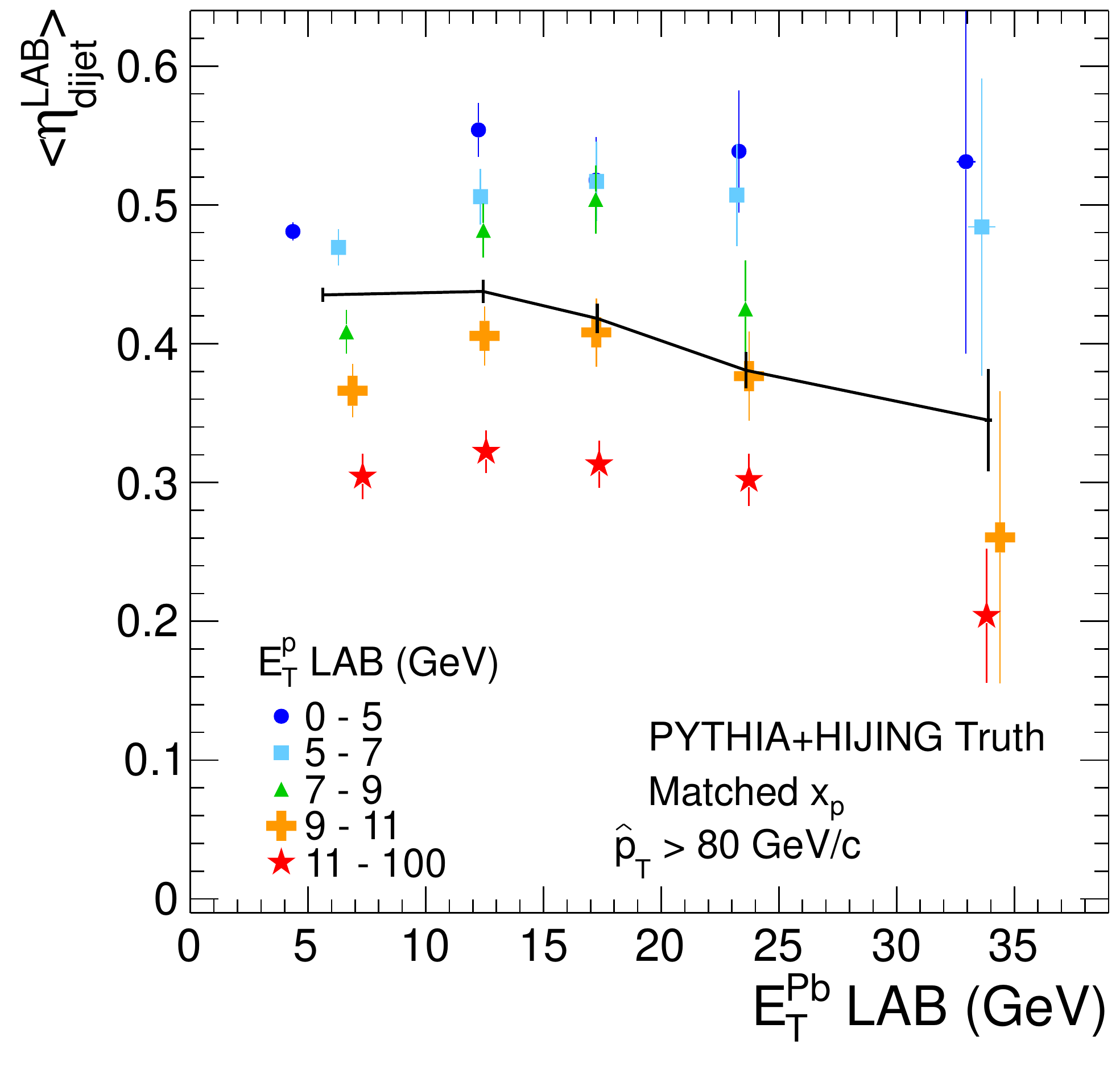}}
\caption{(Left) CMS data \cite{Chatrchyan:2014hqa}, (Right) PYTHIA+HIJING with ${x}_{p}$ matching. Colored markers show average dijet pseudorapidity ($\eta_{\rm dijet}=(\eta_{\rm 1}+\eta_{2})/2$) as a function of transverse energy deposition in the Pb-going direction, $E_{\rm T}^{\rm Pb}$, for bins of the transverse energy deposition in the p-going direction, $E_{\rm T}^{p}$. The choice of pseudorapidity intervals for $E_{\rm T}^{\rm Pb}$ and $E_{\rm T}^{p}$ are $-5 < \eta < -4$ and $4 < \eta < 5$ respectively. The black line shows average dijet pseudorapidity as a function of $E_{\rm T}^{\rm Pb}$ when the requirements on transverse energy in proton going direction are removed.}
\label{mean_dijeteta_fixedETp}
\end{figure}

CMS \cite{Chatrchyan:2014hqa} also provided data on the average dijet pseudorapidity for a fixed energy in the p-going direction, $E_{\rm T}^{p}$ for $4 < \eta < 5$, as a function of the energy in the Pb-going direction, $E_{\rm T}^{\rm Pb}$ for $-5 < \eta < -4$. In figure \ref{mean_dijeteta_fixedETp} we show the comparison of the CMS data with our PYTHIA+HIJING results. Once more, the model captures the trend of data except for the lowest activity data where only one or two nucleons from the Pb nucleus contribute and the model -- that corrects for energy only on the proton side -- is clearly deficient due to the neglect of energy-momentum constraints for Pb.

\section{ATLAS jet results}
\label{atlas}

In this section we consider the single-jet measurements by ATLAS \cite{ATLAS:2014cpa}. We generate hard events in PYTHIA with jets reconstructed using the anti-$k_T$ sequential recombination algorithm \cite{Cacciari:2008gp,Cacciari:2011ma} with a
distance parameter of 0.4, in the region $|\eta_{\rm jet}-\eta_{\rm CM}|<3$. The centrality criterium is, in this case, the total transverse energy in Pb-going direction within the pseudorapidity range of $-4.9 < \eta < -3.2$, thus less sensitive to energy constraints on the proton. Let us note that we use, for the different centrality classes, the number of collisions $N_{coll}$ provided by ATLAS and not the one extracted from HIJING \footnote{They are  (11.94,9.86,8.38,6.934,4.82,2.29) for HIJING and (14.57,12.07,10.37,8.94,6.44,2.98) in the ATLAS paper \cite{ATLAS:2014cpa}, for the $0-10 \%$, $10-20 \%$, $20-30 \%$, $30-40 \%$, $40-60 \%$ and $60-90 \%$ centrality classes respectively.}.

\begin{figure}[!ht]
 \centering
\resizebox{0.8\textwidth}{!}{\includegraphics{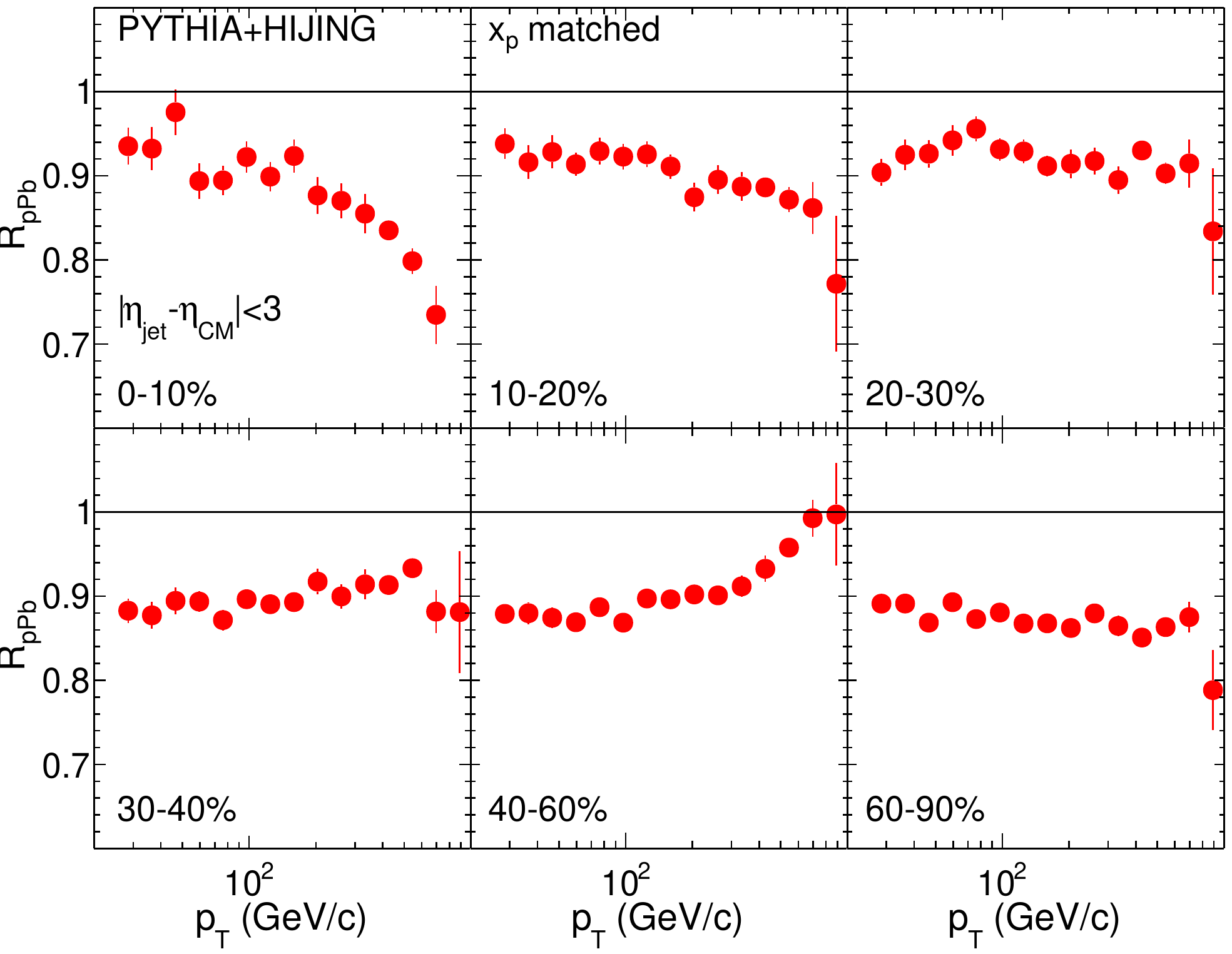}}
\caption{${\rm R}_{p{\rm Pb}}$ calculation in PYTHIA+HIJING with $x_p$ matching are shown for $0-10 \%$, $10-20 \%$, $20-30 \%$, $30-40 \%$, $40-60 \%$ and $60-90 \%$ centrality classes. The centrality classes are determined according to total transverse energy in Pb-going direction within the pseudorapidity range $-4.9 < \eta < -3.2$.}
\label{RpPb_inclusiveeta}
\end{figure}

In figure \ref{RpPb_inclusiveeta} we show the results of the model for the nuclear modification factor ${\rm R}_{p{\rm Pb}}$ of jets as a function of their transverse momentum, for different centrality classes, integrated over the whole pseudorapidity region $|\eta_{\rm jet}-\eta_{\rm CM}|<3$. The effect 
of the centrality selection becomes evident.

\begin{figure}[!ht]
 \centering
\resizebox{0.85\textwidth}{!}{\includegraphics{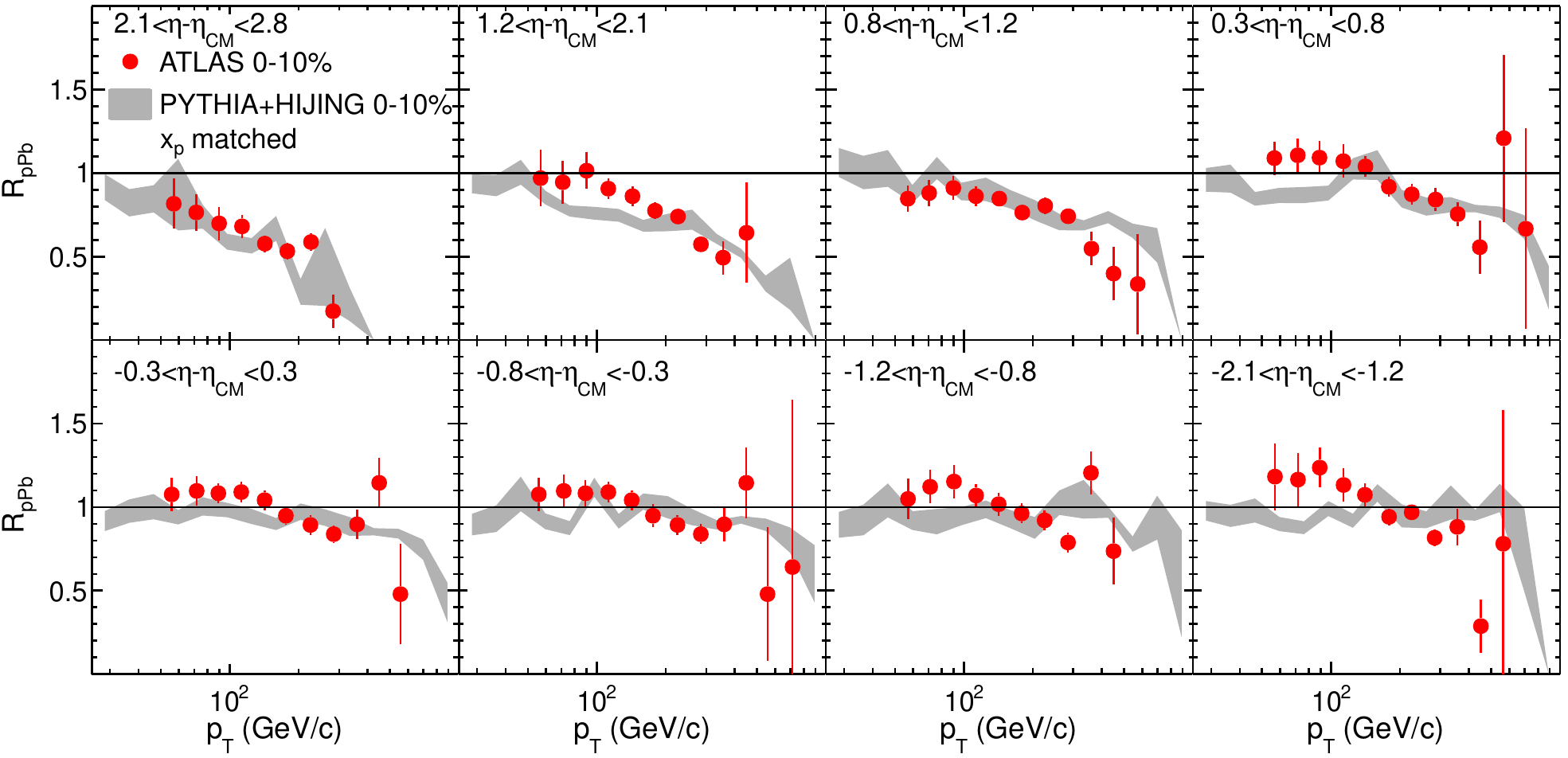}}
\caption{${\rm R}_{p{\rm Pb}}$ calculation in PYTHIA+HIJING with $x_p$ matching (grey bands) and the measured values by ATLAS (\cite{ATLAS:2014cpa}, red circles) are shown for the $0-10 \%$ centrality class in bins of pseudorapidity in the center-of-mass frame. The centrality classes are determined according to total transverse energy in Pb-going direction within the pseudorapidity range $-4.9 < \eta < -3.2$.}
\label{RpPb_central}
\end{figure}

\begin{figure}[!ht]
 \centering
\resizebox{0.85\textwidth}{!}{\includegraphics{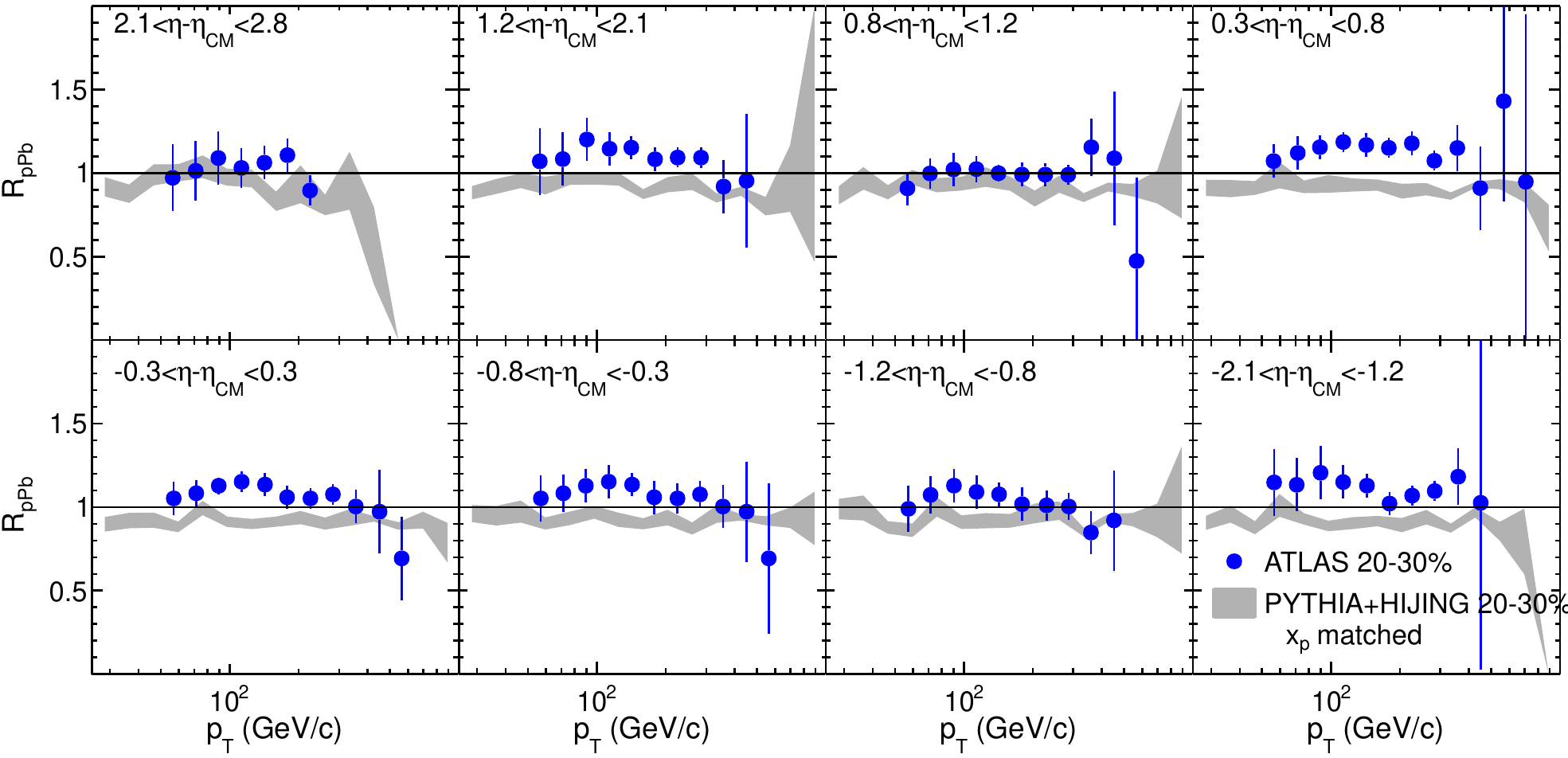}}
\caption{${\rm R}_{p{\rm Pb}}$ calculation in PYTHIA+HIJING with $x_p$ matching (grey bands) and the measured values by ATLAS (\cite{ATLAS:2014cpa}, blue circles) are shown for the $20-30 \%$ centrality class in bins of pseudorapidity in the center-of-mass frame. The centrality classes are determined according to total transverse energy in Pb-going direction within the pseudorapidity range $-4.9 < \eta < -3.2$.}
\label{RpPb_midcentral}
\end{figure}

\begin{figure}[!ht]
 \centering
\resizebox{0.85\textwidth}{!}{\includegraphics{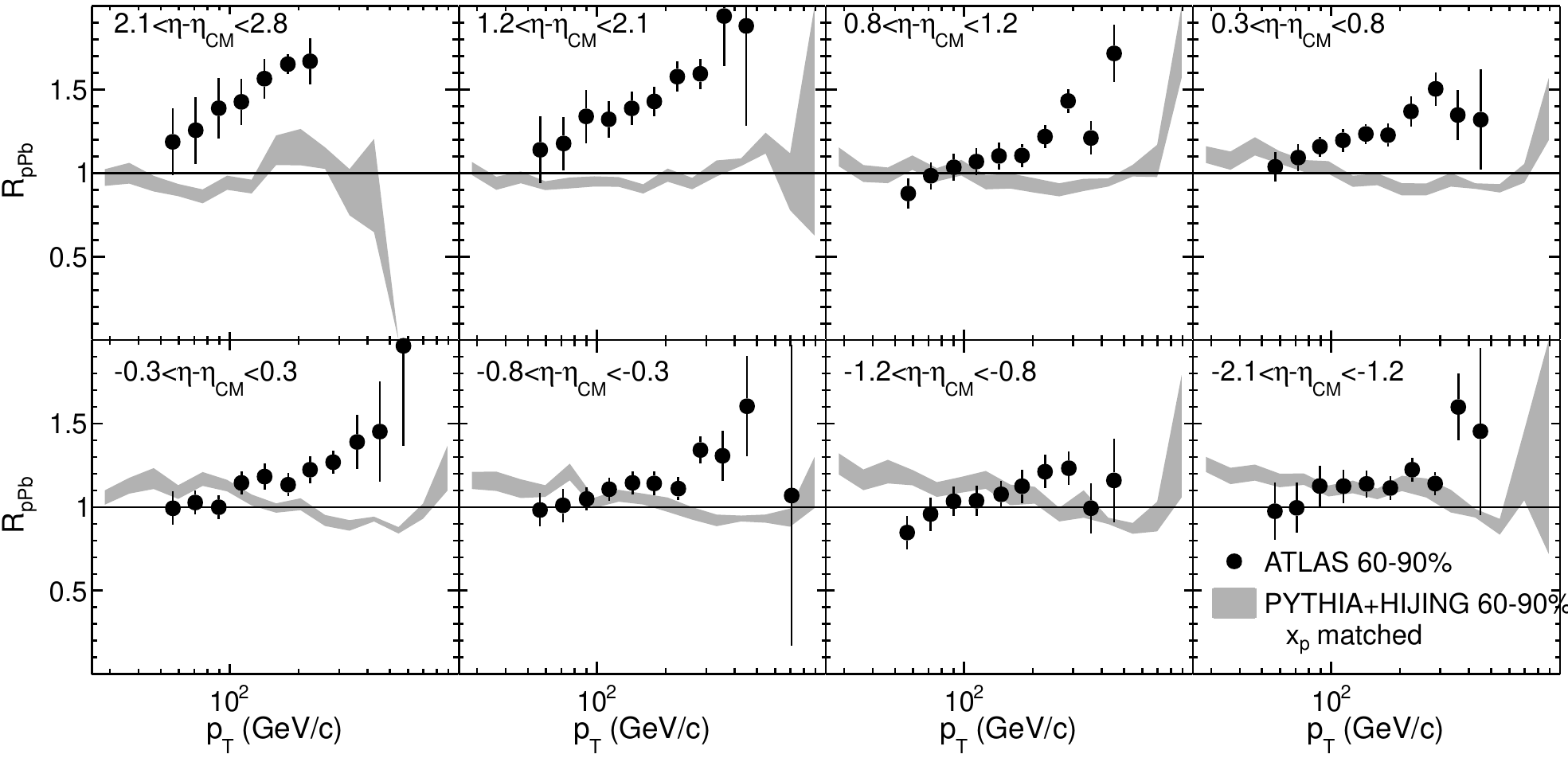}}
\caption{${\rm R}_{p{\rm Pb}}$ calculation in PYTHIA+HIJING with $x_p$ matching (grey bands) and the measured values by ATLAS (\cite{ATLAS:2014cpa}, black circles) are shown for the $60-90 \%$ centrality class in bins of pseudorapidity in the center-of-mass frame. The centrality classes are determined according to total transverse energy in Pb-going direction within the pseudorapidity range $-4.9 < \eta < -3.2$.}
\label{RpPb_peripheral}
\end{figure}

In figures \ref{RpPb_central}, \ref{RpPb_midcentral} and \ref{RpPb_peripheral} we show a comparison of the results of the model with ATLAS data \cite{ATLAS:2014cpa} for the nuclear modification factor ${\rm R}_{p{\rm Pb}}$ of jets as a function of their transverse momentum in different pseudorapidity bins, for central, semicentral and peripheral collisions respectively. A good agreement with data is found for central collisions that deteriorates with decreasing centrality, until the model fails for peripheral collisions as expected from previous discussions.

\section{Conclusions}
\label{conclu}

In this work we have analyzed the effect of centrality selection on events characterised by a hard observable in $p$Pb collisions at the LHC. We have focused on the influence of the interplay given by simple considerations of energy-momentum conservation, between the hard process that is the observable of interest -- jet production -- and the underlying event that provides the centrality estimator. We have developed a simplistic model that considers first the production of jets in a $pp$ collision as described by PYTHIA. From each $pp$ collision, the value of the energy of the parton from the proton participating in the hard scattering is extracted. Then, the underlying event is generated simulating a $p$Pb collision through HIJING, but with the energy of the proton decreased according to the value extracted in the previous step. The full event consists in the superposition of the hard and underlying ones. In this way the energy constraints on the proton are taken into account, while those on the Pb nucleus are not considered. The model is thus expected to fail for peripheral collisions where very few nucleons from Pb participate. Note that our model is not a dynamical one, see other explanations in \cite{Martinez-Garcia:2014ada,Bzdak:2014rca,Alvioli:2014eda,Perepelitsa:2014yta}, our only aim being the study of how important this class of biases may become on existing experimental data.

We have considered two sets of  data: dijets from CMS \cite{Chatrchyan:2014hqa} and single jets from ATLAS \cite{ATLAS:2014cpa} and find that the model is able to capture the bulk of the centrality effect for central to semicentral collisions, while it fails -- as it should -- for peripheral collisions. We conclude that this simple bias due to energy-momentum conservation must be considered before any extraction of other eventual centrality dependencies as those of nuclear parton densities.

The obvious extension of this work is to include the energy-momentum constraints on the nucleus side, which we plan to address. In any case, we find that data as those discussed in this manuscript, offer most valuable information for the development of models for high-energy hadronic and nuclear collisions, as they link perturbative computable quantities as jets with those coming from the underlying event that can only be addressed through accurate modelling soft particle production. The present discussions on centrality definition in asymmetric collisions make it clear that such models are critically needed.

\section*{Acknowledgements}

 We thank Brian Cole, Yen-Jie Lee and Gunther Roland for useful discussions. NA  thanks the Theory
Unit at CERN for hospitality and support during stays when part of this work was developed.
The work of NA is supported by the European Research Council grant HotLHC ERC-2011-
StG-279579; by Ministerio de Ciencia e Innovaci\'on of Spain under projects FPA2011-22776; by Xunta de Galicia (Conseller\'{\i}a de Educaci\'on and Conseller\'{\i}a de Innovaci\'on
e Industria -- Programa Incite); and by the Spanish Consolider-Ingenio 2010
Programme CPAN and by FEDER. The work of DCG is supported by the U.S. Department of Energy under cooperative research agreement DE-SC0011088. The work of JGM is supported by Funda\c c\~ao para a Ci\^encia e
a Tecnologia (Portugal) under project CERN/FP/123596/2011 and contract
Investigador FCT -- Development Grant.
 

 
\end{document}